\documentclass{easychair}

\usepackage{doc}

\usepackage[T1]{fontenc} 
\usepackage{amsmath} 
\usepackage{amsfonts} 
\usepackage{enumitem}\setlist[itemize]{noitemsep,topsep=-\parskip}
\usepackage{listings}
\usepackage{xcolor}
\definecolor{lightgrey}{RGB}{240,240,240}
\lstdefinelanguage{Rocq}
{
  inputencoding=utf8,
  extendedchars=true,
  numbers=none,
  numberstyle={},
  tabsize=2,
  basicstyle={\ttfamily\footnotesize\upshape},
  backgroundcolor=\color{yellow!5},
  keywords={Axiom,Canonical,Definition,end,End,Fixpoint,Lemma,match,Proof,Prop,Qed,Record,Section,Theorem,Type,Variable,Variables,with,Inductive,Ltac,let,in,forall,exists,Tactic,Notation},
  sensitive=true,
  keywordstyle=\color{blue},
  morecomment=[n]{(*}{*)},
  commentstyle={\itshape\color{red}},
  string=[b]{"},
  stringstyle=\color{orange},
  showstringspaces=false,
  literate=
  {λ}{$\lambda$}1
  {↪}{$\hookrightarrow$}1
  {→}{$\rightarrow$}1
  {Π}{$\Pi$}1
  {≔}{$\coloneqq$}1
  {⊢}{$\vdash$}1
  {≡}{$\equiv$}1
  {𝔹}{$\mathbb{B}$}1
  {𝕃}{$\mathbb{L}$}1
  {ℕ}{$\mathbb{N}$}1
  {α}{$\alpha$}1
  {β}{$\beta$}1
  {η}{$\eta$}1
  {π}{$\pi$}1
  {τ}{$\tau$}1
  {ω}{$\omega$}1
  {∧}{$\wedge$}1
  {≤}{$\le$}1
  {≠}{$\neq$}1
  {∉}{$\notin$}1
  {×}{$\times$}1
  {⋅}{$\cdot$}1
  {ε}{$\varepsilon$}1
  {∃}{$\exists$}1
  {¬}{$\neg$}1
  {∨}{$\vee$}1
  {∀}{$\forall$}1
  {⇒}{$\Rightarrow$}1
  {∧}{$\wedge$}1
  {ᵢ}{$_i$}1
  {ₑ}{$_e$}1
  {₁}{$_1$}1
  {₂}{$_2$}1
}

\usepackage{mathtools} 
\usepackage{latexsym} 
\lstdefinelanguage{Lambdapi}
{
  inputencoding=utf8,
  extendedchars=true,
  numbers=none,
  numberstyle={},
  tabsize=2,
  basicstyle={\ttfamily\small\upshape},
  backgroundcolor=\color{blue!5},
  keywords={abort,admit,admitted,apply,as,assert,assertnot,associative,assume,begin,builtin,commutative,compute,constant,debug,end,fail,flag,focus,generalize,have,in,induction,inductive,infix,injective,left,let,notation,off,on,opaque,open,prefix,print,private,proofterm,protected,prover,prover_timeout,quantifier,refine,reflexivity,require,rewrite,right,rule,sequential,simplify,solve,symbol,symmetry,type,TYPE,unif_rule,verbose,why3,with},
  sensitive=true,
  keywordstyle=\color{blue},
  morecomment=[l]{//},
  morecomment=[n]{/*}{*/},
  commentstyle={\itshape\color{red}},
  string=[b]{"},
  stringstyle=\color{orange},
  showstringspaces=false,
  literate=
  {λ}{$\lambda$}1
  {↪}{$\hookrightarrow$}1
  {→}{$\rightarrow$}1
  {Π}{$\Pi$}1
  {≔}{$\coloneqq$}1
  {⊢}{$\vdash$}1
  {≡}{$\equiv$}1
  {𝔹}{$\mathbb{B}$}1
  {𝕃}{$\mathbb{L}$}1
  {ℕ}{$\mathbb{N}$}1
  {α}{$\alpha$}1
  {β}{$\beta$}1
  {η}{$\eta$}1
  {π}{$\pi$}1
  {τ}{$\tau$}1
  {ω}{$\omega$}1
  {∧}{$\wedge$}1
  {≤}{$\le$}1
  {≠}{$\neq$}1
  {∉}{$\notin$}1
  {×}{$\times$}1
  {⋅}{$\cdot$}1
  {ε}{$\varepsilon$}1
  {∃}{$\exists$}1
  {¬}{$\neg$}1
  {∨}{$\vee$}1
  {∀}{$\forall$}1
  {⇒}{$\Rightarrow$}1
  {∧}{$\wedge$}1
  {ᵢ}{$_i$}1
  {ₑ}{$_e$}1
  {₁}{$_1$}1
  {₂}{$_2$}1
}

\newcommand\hide[1]{}

\newcommand\eg{{\em e.g.} }
\newcommand\ie{{\em i.e.} }

\newcommand\furl[1]{\footnote{\url{http://#1}}}

\renewcommand\prod{\mr{prod}}

\renewcommand\a{\rightarrow}
\newcommand\A{\Rightarrow}

\newcommand\ra\hookrightarrow 

\newcommand\ex{\exists}
\newcommand\all{\forall}
\newcommand\ou{\vee}

\newcommand\et{\wedge}

\renewcommand\th{\vdash}

\newcommand\al{\alpha}
\renewcommand\b{\beta}

\newcommand\vep{\varepsilon}

\renewcommand\l{\lambda}

\newcommand\mt{\mathtt}
\newcommand\mr{\mathrm}
\newcommand\mb{\mathbb}

\newcommand\bN{\mb{N}}

\newcommand\bR{\mb{R}}

\newenvironment{rul}
  {$\begin{array}{rcl}}
  {\end{array}$}

\newenvironment{rew}[1][~~\a~~]
  {$\begin{array}{r@{#1}l}}
  {\end{array}$}
\newenvironment{rewc}[1][~~\a~~]
  {\begin{center}\begin{rew}[#1]}
  {\end{rew}\end{center}}

\newenvironment{lstgeneric}[2]
  {\begin{list}{#1}{\topsep=.5mm\itemsep=.5mm\parsep=0mm%
    \itemindent=-3ex\labelsep=1ex\labelwidth=0ex #2}}
  {\end{list}}

\newcommand\merci{The authors thank Anthony Bordg for his
  contributions on the alignments of the types {\tt option}, {\tt sum}
  and {\tt list}, Amal Makni for her contributions on the alignment of
  subtypes, quotient types and real numbers, Théo Winterhalter for his
  contributions on the alignment of integer functions, Alessio
  Coltellacci for his contribution on the alignment of the {\tt lcm}
  function, Jérémy Dubut for his contributions on the alignment of
  some predicates, Reynald Affeldt for his help with Ssreflect and
  Mathcomp, and Catherine Dubois for her comments on a previous
  version of this paper.}

\renewcommand\tilde{\raise.17ex\hbox{$\scriptstyle\sim$}}

\title{Aligning HOL-Light and Rocq libraries formally\thanks\merci}

\author{\href{https://blanqui.gitlabpages.inria.fr/}{Fr\'ed\'eric Blanqui}\inst{1}$^b$ \and \href{https://agontard.github.io/}{Antoine Gontard}\inst{1}$^a$}

\institute{{\bf $^a$Universit\'e Paris-Saclay}, {\bf $^b$INRIA}, ENS Paris-Saclay, CNRS\\Laboratoire M\'ethodes Formelles, 4 av des Sciences, 91190 Gif-sur-Yvette, France}

\authorrunning{F. Blanqui and A. Gontard}

\titlerunning{Aligning HOL-Light and Rocq libraries formally}

\begin{document}

\maketitle

\begin{abstract}
  We report on our efforts to translate HOL-Light theorems on HOL-Light
types/functions into Rocq theorems on Rocq types/functions. To this
end, we developed in Rocq tactics to automate the proofs required for
replacing a HOL-Light inductive type or recursive function definition
by an equivalent but more idiomatic one in Rocq. We also explain how
we replaced the definition of real numbers, as well as a number of
mathematical notions like $\mathbb{R}^n$ spaces and the definition of
limit, hence providing to Rocq users many definitions and theorems in
logic and analysis that had not been formalized in Rocq before.


\end{abstract}



%
%

Texts in blue are clickable hyperlinks.

\section{Introduction}
There exist many different proof systems, each one with its pros and
cons depending on the targeted application. Each system comes with
libraries containing definitions and proofs of theorems. Some
definitions and theorems are available in all systems, \eg the
definition of addition on natural numbers and the proofs of its basic
properties. But there are also definitions and theorems that are
available in one system
only\footnote{\url{https://www.cs.ru.nl/~freek/100/}} because it took
several man-years to be formalized, \eg the Feit-Thompson odd-order
theorem is available in Rocq only \cite{gonthier13itp}, Green's
theorem is available in Isabelle/HOL only \cite{abdulaziz19jar}, the
definition of Scholze's perfectoid spaces is available in Lean only
\cite{buzzard20cpp}, the proof that there are exactly 5 Platonic solids is
available in HOL-Light only, \ldots\ Unfortunately there exist few tools
that can translate this kind of definitions and proofs to other systems
\cite{kohlhase21jar}.

One difficulty comes from the fact that proof systems are often based
on different logics, sometimes incompatible. In this paper, we consider
the special case of HOL-Light and Rocq but our methodology and tactics
could be adapted to other source systems that are, like HOL-Light, based on
higher-order logic with choice (\eg HOL4, Isabelle/HOL), and other target
systems that are, like Rocq, based on the calculus of inductive constructions
(\eg Agda, Lean).

To ease the translation between two different systems, it is
possible to represent their deduction rules and proofs in a logical
framework \cite{pfenning01chapter}. Using a logical framework instead of
developing one-to-one translations allows one to more easily add new
systems as sources or as targets independently. Several logical
frameworks have been developed over the years, \eg predicate calculus,
Automath \hide{\cite{debruijn91chapter,}}\cite{geuvers13im}, LF
\cite{harper93jacm,pfenning99cade}\hide{pfenning94cade}, Martin-Löf's
logical framework \cite{martinlof84book,magnusson93types},
$\lambda$Prolog \cite{miller91jlc}, Isabelle
\cite{paulson94lncs}\hide{paulson89jar}.

In this work, we will use the
\href{https://github.com/Deducteam/Dedukti}{Dedukti} logical framework
\cite{assaf16draft} as it is actively used, with many tools translating
proofs from various systems to Dedukti, \eg
\href{https://github.com/Deducteam/hol2dk}{Hol2dk} for HOL-Light
\cite{blanqui24lpar},
\href{https://github.com/Deducteam/CoqInE}{CoqInE} for Rocq
\cite{boespflug12pxtp-coqine},
\href{https://github.com/Deducteam/lean2dk}{Lean2dk} for Lean,
\href{https://github.com/Deducteam/Agda2Dedukti}{Agda2Dedukti} for
Agda, and from Dedukti to various systems, \eg
\href{https://github.com/Deducteam/sttfaxport}{STTfaXport} for Matita,
PVS, OpenTheory \cite{thire18lfmtp},
\href{https://github.com/Deducteam/lambdapi}{Lambdapi} for Rocq
\cite{blanqui24lpar},
\href{https://github.com/Deducteam/predicativize}{Predicativize} for
Agda \cite{felicissimo24lmcs}.

Dedukti is based on the $\lambda\Pi$-calculus modulo rewriting, an
extension of LF with implicit identification of types modulo
user-defined oriented equations \cite{assaf16draft}. It allows more
shallow embeddings than LF and to reduce the size of proofs
further. The propositional equality of two
$\b$-equivalent terms requires a long proof in HOL-Light and only a
one-step proof in LF and Dedukti (an application of reflexivity) because
$\b$-equivalent terms are identified in them. It has also been proved
that the deduction rules and proofs of many different proof systems,
from (fragments) of first or higher-order logic to extensions of the
calculus of constructions, can be represented in Dedukti in such a way
that a feature (\eg polymorphism, impredicativity, dependent types)
common to two different systems can be represented by the same
construction in Dedukti, which helps in comparing and translating
proofs between different systems \cite{blanqui23lmcs}.

However, translating the definitions and proofs from one system to
another is not enough. To get directly usable theorems in the target
system, it is necessary to replace in the translated theorems the
types and functions of the source system with those already defined in
the target system \cite{kohlhase21jar}. For instance, a Rocq user
willing to reuse a HOL-Light theorem on real numbers does not want
such a translation to provide them with a Rocq theorem on the
representation of HOL-Light reals in Rocq, but with a Rocq theorem on
Rocq reals as they are already defined in the Rocq
\href{https://rocq-prover.org/doc/V9.1.0/stdlib/index.html}{standard}
or \href{https://github.com/math-comp/math-comp}{Mathcomp}
libraries. However, for such a replacement to be valid, it is
necessary to prove that the types and functions of the target system
satisfy the properties characterizing them in the source system. We
then say that those types and functions are formally aligned
\cite{kaliszyk16cicm-wip}.

In \cite{blanqui24lpar}, the first author introduced a method and the
tools \href{https://github.com/Deducteam/hol2dk}{Hol2dk} and
\href{https://github.com/Deducteam/lambdapi}{Lambdapi} to extract
definitions and proofs from HOL-Light\hide{\footnote{Since HOL-Light is
implemented with the LCF approach \cite{paulson87book}, it needs to be
instrumented to record the deduction rule used at each step.}} and
translate them to Rocq by using Dedukti as intermediate language. He
also sketched how to obtain a few hundred theorems of arithmetic
directly usable in Rocq by aligning by hand a few basic types and
functions on natural numbers.

In this paper, we present a number of Rocq tactics that we developed
to automate the proofs of the properties required to formally align
the HOL-Light definition of any inductive type or primitive recursive
function with its Rocq idiomatic counterpart. We applied those tactics
to automatically translate to Rocq the HOL-Light
\href{https://github.com/jrh13/hol-light/blob/master/Logic/make.ml}{Logic}
library, providing Rocq users with a number of directly usable
theorems that had not been formalized in Rocq before (\eg
Skolem-Gödel-Herbrand, Herbrand and Birkhoff theorems).

We also explain how we formally aligned the type of real numbers
between both systems, as well as a number of mathematical objects and
notions like the sets $\bR^n$ and the notion of limit. We used those
new alignments to translate to Rocq the HOL-Light
\href{https://github.com/jrh13/hol-light/blob/master/Multivariate/make_complex.ml}{Multivariate}
library, which contains more than 20,000 theorems in real and complex
analysis, topology, \ldots\ hence providing to Rocq users theorems that
had not been formalized in Rocq before, like the ratio test theorem
for the convergence of series.

All these theorems are gathered in the Opam package
\href{https://github.com/Deducteam/rocq-hollight}{rocq-hollight}.

We start by introducing the two systems and their features which are
relevant to this work. We then explain how to translate HOL-Light
definitions and proofs to Rocq and replace in Rocq a type or function
defined {\it à la} HOL-Light by a type or function defined in a more
idiomatic way. We then detail our automation tactics before presenting
the translated Rocq libraries. In the conclusion, we discuss how this
work can be adapted for translations between other source and target
systems, and quickly address additional challenges if one wants to do
the reverse translation from Rocq to HOL-Light.


\vspace*{2mm}
\noindent\href{https://rocq-prover.org/}{\bf Rocq} is a proof
assistant based on the calculus of inductive constructions, an
extension of the calculus of constructions with primitives for
inductive type and primitive recursive function definitions
\cite{paulin93tlca}. Rocq provides a command {\tt Inductive} to define
an inductive type and generate its associated induction principle, and
a command {\tt Fixpoint} to define primitive recursive functions. For instance,
the inductive type of polymorphic lists and the addition on (unary)
Peano natural numbers are defined in the Rocq Core library as
follows:\footnote{Code with a yellow background is Rocq code, while
code with a blue background is Lambdapi code, used for describing
mappings from HOL-Light to Rocq.}

\begin{lstlisting}[language=Rocq]
Inductive list (A : Type) :=
  nil : list A | cons : A -> list A -> list A.

Fixpoint add n m :=
  match n with 0 => m | S p => S (add p m) end.
\end{lstlisting}

The calculus of constructions \cite{coquand88ic} is an extension of
simple type theory with dependent and polymorphic types, that is,
terms and types can take terms and types as arguments. The expression
{\tt forall x:A,B} is used to represent the type of functions mapping
a term {\tt x} of type {\tt A} to a term of type {\tt B}, where {\tt
  B} can depend on {\tt x}. It boils down to the simple function type
{\tt A->B} if {\tt B} does not depend on {\tt x}. Moreover, in Rocq,
proofs are themselves terms, that is, terms and types can take proofs
as arguments. Hence, as we shall see in the next section, one can
easily define a general notion of subtype in Rocq, which will be
useful to translate HOL-Light types to Rocq.

Based on the proposition-as-type approach, proving a proposition $P$
in Rocq consists in building a term of type $P$. This
yet-to-be-defined term is implemented in Rocq by a metavariable that
needs to be instantiated. To this end, Rocq provides primitive tactics
as well as a tactic language for defining new tactics
\cite{delahaye00lpar}. In its simplest form, a tactic has the effect
of instantiating a metavariable representing a goal by some term
possibly containing new metavariables/subgoals. For instance, from a
goal of the form {\tt forall n:nat, P n}, the {\tt induction} tactic
uses the automatically generated induction principle for the type {\tt
  nat} to create two new subgoals, one for {\tt P 0} and one for {\tt
  forall n:nat,P n -> P (S n)}. A theorem is proved only when all
metavariables have been instantiated.


\vspace*{2mm}\noindent{\href{https://github.com/jrh13/hol-light/}{\bf HOL-Light}}
is a proof assistant based on simple type theory with prenex
polymorphism and, at the beginning, only two type constants: $\mt{bool}$
for propositions and $\mt{ind}$ for individuals. It implements a
variant of Andrews' Q0 logic \cite{andrews02book} where everything,
including connectives, $\alpha$-equivalence, $\beta$-equivalence, the
introduction and elimination rules of logical connectives, etc., is
defined from the equality symbol ${=}:{\alpha\a\alpha\a\mt{bool}}$
(where $\alpha$ is a type variable implicitly universally quantified),
a choice operator $\varepsilon:(\alpha\a\mt{bool})\a\alpha$, the
axioms $\th\all t,(\l x,(t~x))=t$ and $\th\all
P:\alpha\a\mt{bool},\all x:\alpha,P~x\a P~(\varepsilon~P)$, and a few
more deduction rules \cite{harrison09tphol}. It also assumes a
function $\mt{IND\_SUC}:\mt{ind}\a\mt{ind}$ that is injective but not
surjective, implying that $\mt{ind}$ is infinite.

As a consequence, in HOL-Light, every proposition is either true or
false (excluded middle axiom of classical logic), two propositions are
equal if they are logically equivalent (propositional extensionality),
and two functions are equal if they are pointwise equal (functional
extensionality).

\noindent
Everything else is built step by step by using two additional
mechanisms:
\begin{itemize}
\item With the command
  \href{https://hol-light.github.io/references/HTML/new_definition.html}{\tt
    new\_definition}, a user can introduce a new function symbol $f$
  with an axiom of the form $\th f=t$ where $t$ is any term not
  containing $f$. For instance, the logical connectives are defined in
  this way: $\th{\top}={((\l p,p)=(\l p,p))}$, $\th{\all}={(\l p,(p=\l
    x,\top))}$, \ldots

\item Given an already defined type constant $A$, a predicate
  $P:A\a\mt{bool}$ and a proof of $\th\ex x\!:\!A,P x$, a user can
  introduce with \href{https://hol-light.github.io/references/HTML/new_type_definition.html}{\tt new\_type\_definition}:
\begin{itemize}
\item a new type constant $B$ together with:
\item two new term constants $mk\!:\!A\a B$ and $dest\!:\!B\a A$, and
  two new axioms:
\item $\th\all b\!:\!B,mk(dest~b)=b$, stating that $mk$ is surjective and
  $dest$ injective,
\item $\th\all a\!:\!A,(P~a)=(dest(mk~a)=a)$, stating that the image
  of $dest$ is the set of the elements of $A$ satisfying $P$, and that
  $mk$ is injective on it.
\end{itemize}
\end{itemize}

\noindent
For instance, the HOL-Light base library axiomatizes the polymorphic
binary product type constructor $\mt{prod}~\al~\b$ as an isomorphic
image of the subset of $\al\a\b\a\mt{bool}$ of the functions $f$
satisfying the predicate $P$ such that
${P(f)}={(\ex a:\al,\ex b:\b,{{f}={(\lambda x:\al,\lambda y:\b,x=a\et y=b)}})}$.
It also axiomatizes the type $\mt{num}$ of natural
numbers as an isomorphic image of the smallest subset of $\mt{ind}$
stable by $\mt{IND\_SUC}$ and containing a chosen element
$\mt{IND\_0}$ that is not in the image of $\mt{IND\_SUC}$. The natural
number $0$ is then defined as $mk(\mt{IND\_0})$, and the successor
function $\mt{SUC}:\mt{num}\a\mt{num}$ as $\lambda
n:\mt{num},mk(\mt{IND\_SUC}(dest~n))$.


    \vspace*{1mm}\noindent{\bf Definition of inductive types in HOL-Light.}
Every inductive type is defined by the command
\href{https://www.cl.cam.ac.uk/~jrh13/hol-light/HTML/define_type.html}{\tt
  define\_type} as a subset of an instance of the polymorphic type
$\mt{recspace}~\al$
\cite{melham89chapter,gunter93tphol,harrison95tphol}. 
The type $\mt{recspace}~\al$ itself is defined as a subset of
$\mt{num}\a\al\a\mt{bool}$ and resembles Martin-L\"of's type $W$
\cite{martinlof84book}. It has two constructors:
$\mt{BOTTOM}:\mt{recspace}~\alpha$ and
$\mt{CONSTR}:\mt{num}\a\alpha\a(\mt{num}\a\mt{recspace}~\alpha)\a\mt{recspace}~\alpha$,
which is used to encode inductive type constructors as follows:
\begin{itemize}
\item The first argument of $\mt{CONSTR}$ is a number
  uniquely identifying the constructor.
\item The second argument is a tuple containing all the non-recursive
  arguments of the constructor. To this end, $\alpha$ is chosen to be
  the product of the types of the non-recursive arguments of all the
  constructors.
\item The last argument encodes the recursive arguments. Although it
  has the type of an infinite sequence, only a finite number of
  recursive arguments is actually allowed. This is enforced by using
  two construction functions, $\mt{FNIL}$ and $\mt{FCONS}$, mimicking
  the usual list constructors\footnote{This is done like this because
  the inductive type of lists is not yet defined at this point in
  HOL-light: it will be later defined as a subset of
  $\mt{recspace}$.}.
\end{itemize}
A user-defined inductive type is then declared as being in bijection
with the subset of $\mt{recspace}~A$ whose elements can be interpreted
as constructors (where $A$ is the product of the types of
non-recursive arguments of constructors).

For instance, for the inductive type $\mt{list}~\alpha$ whose
constructors are $\mt{NIL}:\mt{list}~\alpha$ and
$\mt{CONS}:\alpha\a\mt{list}~\alpha\a\mt{list}~\alpha$:
\begin{itemize}
\item the type $\mt{list}~\al$ is defined as being in bijection with
  the elements of $\mt{recspace}~\al$ satisfying the predicate
  $P_\mt{list}$ (defined hereafter) through some functions\\
  $\mt{\_mk\_list}:\mt{recspace}~\al\a\mt{list}~\al$ and
  $\mt{\_dest\_list}:\mt{list}~\al\a\mt{recspace}~\al$;
\item $\mt{NIL}$ is defined as
  $\mt{\_mk\_list}(\mt{CONSTR}~0~d~\mt{FNIL})$ where
  $d$ is some default value;
\item $\mt{CONS}$ is defined as\\ $\lambda a:\alpha, \lambda
  l:\mt{list}~\alpha,\mt{\_mk\_list}(\mt{CONSTR}~1~a~(\mt{FCONS}~(\mt{\_dest\_list}~l)~\mt{FNIL}))$;
\item $P_\mt{list}$ is the inductively defined predicate/smallest subset of
  $\mt{recspace}~\al$ containing the terms of the form $\mt{CONSTR}~0~d~\mt{FNIL}$ or
  $\mt{CONSTR}~1~a~(\mt{FCONS}~r~\mt{FNIL})$ with $r$ itself in $P_\mt{list}$.
\end{itemize}

\noindent
HOL-Light accepts first-order datatypes only (constructors cannot take
functions as arguments), a strict subset of types accepted by the
Rocq
\href{https://rocq-prover.org/doc/V9.2.0/refman/language/core/inductive.html}{Inductive}
command.

\hide{
We note that any definition accepted by the
\href{https://www.cl.cam.ac.uk/~jrh13/hol-light/HTML/define_type.html}{\tt new\_type}
command would also be accepted by Rocq's
\href{https://rocq-prover.org/doc/V9.2.0/refman/language/core/inductive.html}{\tt Inductive}
command if written with the according syntax, so it is always possible to
define a Rocq-idiomatic equivalent to a HOL-Light inductive type.
}


    \vspace*{1mm}\noindent{\bf Definition of recursive functions in HOL-Light.}
%
Recursive functions are defined using the choice operator
$\varepsilon$, the equations they satisfy and a termination proof
\cite{slind96tphol}. For example, the addition
$\mt{ADD}:\mt{num}\a\mt{num}\a\mt{num}$ is defined as:
\begin{center}
\begin{minipage}{12cm}
$\varepsilon~(\lambda add':\mt{num}\a\mt{num}\a\mt{num}\a\mt{num},\forall a :\mt{num},\\
  \hspace*{12mm}(\forall n :\mt{num},add'~a~(\mt{NUMERAL}~0)~n = n)\\
  \hspace*{7mm}\wedge~(\forall m:\mt{num}, \forall n:\mt{num},
               (add'~a~(\mt{SUC}~m)~n) = (\mt{SUC}~(add'~a~m~n))))\\
  (\mt{NUMERAL}~(\mt{BIT1}~(\mt{BIT1}~(\mt{BIT0}~(\mt{BIT1}~(\mt{BIT0}~(\mt{BIT1}~0)))))))$
\end{minipage}
\end{center}
\noindent
that is, the application of some function $add':\bN^3\a\bN$ satisfying
the equations $add'~a~0~n=n$ and
$add'~a~(\mt{SUC}~m)~n=\mt{SUC}~(add'~a~m~n)$, to a tuple $a$ made of
the ASCII codes of the characters of the function name (this argument
is used to distinguish two functions with the same specification but
different names). Here,
$a=\mt{BIT1}~(\mt{BIT1}~(\mt{BIT0}~(\mt{BIT1}~(\mt{BIT0}~(\mt{BIT1}~0)))))$
is the binary representation of the ASCII code of the character ``+''
(43).

For a survey on recursive function definitions in proof assistants,
see \cite{bove16mscs}.


\section{Translating HOL-Light files to Rocq files}
\label{sec-trans-gen}

Following \cite{blanqui24lpar}, to translate definitions and proofs
from HOL-Light to Rocq, we proceed in two steps. First, with
\href{https://github.com/Deducteam/hol2dk}{Hol2dk} we extract the
definitions and proofs out of HOL-Light and translate them to
Dedukti. Second, with
\href{https://github.com/Deducteam/hol2dk}{Lambdapi}, we translate the
obtained Dedukti files to Rocq files by specifying:
\begin{itemize}\itemsep=0mm
\item a file {\tt encoding.lp} declaring the symbols
  used to encode higher-order logic;
\item a file {\tt renaming.lp} mapping identifiers invalid in Rocq to valid ones;
\item a file {\tt mapping.lp} mapping some Dedukti identifiers to Rocq
  expressions;
\item a list of modules to be imported at the
  beginning of each generated Rocq file.
\end{itemize}
When translating Dedukti files to Rocq, Lambdapi applies the
replacements described in {\tt renaming.lp} and {\tt mapping.lp},
and removes the declarations of the identifiers that are mapped to Rocq
expressions in {\tt mapping.lp}.

By using this mechanism, one can for instance remove the declaration
of the identifier {\tt num} used in HOL-Light for the type of natural
numbers, and replace every occurrence of {\tt num} by the Rocq
expression {\tt nat} which is the Rocq identifier used in the Rocq
standard library for denoting the type of natural numbers. Hence, a
HOL-Light theorem on HOL-Light natural numbers gets translated into a
Rocq theorem on Rocq natural numbers.
However, for this replacement to be valid, one needs to prove that all
the defining properties of {\tt num} are indeed satisfied by {\tt
  nat}.

The translation in itself is not conceptually difficult, since HOL-Light
has a simpler logic than Rocq. The challenge comes from alignments
(which we discuss in the next sections) and having to deal
with huge files. Indeed, because HOL-Light logic is very low-level
(even $\alpha$-equivalence steps are explicit), its proofs get very
big once translated to Rocq.

In order for Rocq to be able to check the proofs generated by the
translation of HOL-Light's {\tt Multivariate} library, we had to
implement many improvements wrt \cite{blanqui24lpar}: split big proofs
into several files, globally share type abbreviations, replace dependencies
on proofs by dependencies on statements, and increase parallelization. We
reduced the checking time by Rocq of the 1 Gb of proofs of the
HOL-Light base library {\tt hol\_lib.ml} from 30 down to 10 minutes
(HOL-Light needs 1m10s to check those proofs). And while the 91 Gb of
proofs of the {\tt Multivariate} library could not be checked in
\cite{blanqui24lpar}, we can now check them in 21 hours\footnote{On an
Intel Core i9-13950HX with 32 threads and 128 Gb RAM, HOL-Light takes
2h30 to check and record the proofs of Multivariate, generating 135 Gb
of proofs which are translated in 1h09 into 91 Gb of Rocq files.}.
  

\section{Methods to align HOL-Light and Rocq definitions}

As explained in the introduction, in HOL-Light, every user-defined
type $A$ or constant/function $f$, is axiomatized by a number of
equations, and all further theorems about $A$ and $f$ only rely on
those equations. Hence, a type $A$ or constant/function $f$ can be
safely replaced by any other type $A'$ or constant/ function $f'$ as
soon as $A'$ and $f'$ satisfy the same equations (this is called an
alignment in \cite{kaliszyk16cicm-wip}).

We first need to replace every type, function or axiom
used to represent HOL-Light's logic in Lambdapi:

\vspace*{1mm}\noindent{\bf The type of HOL-Light types.} The HOL-Light
axiom for $\varepsilon$ implies that any HOL-Light type is
inhabited. Therefore, to translate HOL-Light polymorphism, that is
quantification over types, we need to have a Rocq type for non-empty
types like the following record type made of a carrier type and an
element of this carrier:

\begin{lstlisting}[language=Rocq]
Record Type' := { type :> Type; el : type }.
\end{lstlisting}

\noindent
or the $\mt{pointedType}$ structure of the Rocq Mathcomp library
\cite{mahboubi22book} which has a similar definition. The symbol
$\verb+:>+$ declares $\mt{type}$ as a coercion that is automatically
inserted whenever a term of type $\verb+Type'+$ is encountered at a
position where a term of type $\mt{Type}$ is expected.

\vspace*{1mm}\noindent{\bf HOL-Light native types and equality.}
We then map the HOL-Light type $\mt{ind}$ to the Rocq type {\tt nat}
(this choice is unimportant as $\mt{ind}$ is only used to build
natural numbers; any other infinite type would be fine). We map the
HOL-Light type $\mt{bool}$ to the Rocq type {\tt Prop} of propositions
(and not the Rocq type {\tt bool} of Booleans) because Rocq logical
connectives and equality are defined in {\tt Prop}. We can then map
HOL-Light equality to the equality predicate of the Rocq standard
library. We do not map $\mt{bool}$ to the new Rocq type {\tt SProp} of
proof-irrelevant propositions \cite{gilbert19popl} because it is not
used in the Rocq standard or MathComp libraries (but this could be
useful in a translation to Lean).

\vspace*{1mm}\noindent{\bf HOL-Light axioms and choice operator.}
The remaining HOL-Light axioms and deduction rules are derivable in
Rocq from the following three axioms of the Rocq Mathcomp-analysis
module
\href{https://github.com/math-comp/analysis/blob/master/classical/boolp.v}{boolp}:
functional extensionality, propositional extensionality, and
constructive indefinite description, from which we can define
$\varepsilon$. All together, these axioms imply the excluded middle
\cite{diaconescu75pams}. These are the only axioms required by our
translation.

Although we are not the first to state and use these axioms, we wish
to mention that it is not trivial that they are consistent with Rocq's
logic. As a matter of fact, they were proven to be inconsistent with
older versions of Rocq in \cite{geuvers07note}. Being able to use
these axioms was one of the main motivations for a change in Rocq from
version 8.0 on preventing the proof in \cite{geuvers07note} to work,
as discussed on Rocq's
\href{https://github.com/rocq-prover/rocq/wiki/Impredicative-Set}{wiki}
and reference
\href{https://rocq-prover.org/doc/master/refman/changes.html#details-of-changes-in-8-0}{manual}.

\vspace*{1mm}\noindent{\bf Conditions to align a user-defined type.}
In the case of a user type definition, we need to define the functions
$mk$ and $dest$ and prove that they form a bijection. For instance,
with no entry for {\tt num} and its associated functions and axioms in
{\tt mapping.lp}, Lambdapi generates:
\begin{lstlisting}[language=Rocq]
Axiom num : Type'.
Axiom mk_num : ind -> num.
Axiom dest_num : num -> ind.
Axiom axiom_7 : forall (a : num), (mk_num (dest_num a)) = a.
Axiom axiom_8:
  forall (r:ind), (NUM_REP r) = ((dest_num (mk_num r)) = r).
\end{lstlisting}
which is the representation in Rocq of the HOL-Light definition of
$\mt{num}$. The only way to get rid of all these axioms is to provide
definitions for {\tt num}, {\tt mk\_num} and {\tt dest\_num}, and
proofs for {\tt axiom\_7} and {\tt axiom\_8}, and then redo the
translation by adding in {\tt mapping.lp} mappings for these
constants:

\begin{lstlisting}[language=Lambdapi]
builtin "nat" ≔ num;
builtin "mk_num" ≔ mk_num;
builtin "dest_num" ≔ dest_num;
builtin "lemma_7" ≔ axiom_7;
builtin "lemma_8" ≔ axiom_8;
\end{lstlisting}

\noindent
where mappings are read from right to left: the HOL-Light symbol {\tt
  num} is replaced by the Rocq symbol {\tt nat} which is defined in
the Rocq standard library, the HOL-Light symbol {\tt mk\_num} is
replaced by the Rocq symbol {\tt mk\_num} that we defined ourselves,
etc. (we use the same name whenever possible to help tracing back
symbols from Rocq to HOL-Light).

\vspace*{1mm}\noindent{\bf Default alignment of user-defined types.}
Because in Rocq, proofs are terms and types can depend on terms, it is
always possible to eliminate the axioms introduced by HOL-Light for
defining new types by mapping them to the following generic subtype
construction in Rocq:

\begin{lstlisting}[language=Rocq]
Section Subtype.
  Variables (A : Type) (P : A -> Prop) (a : A) (h : P a).
  Definition subtype := {|type:={x:A|P x}; el:=exist P a h|}.
  Definition dest : subtype -> A := fun x => proj1_sig x.
  Definition mk : A -> subtype := fun x =>
    COND_dep (P x) subtype (exist P x) (fun _ => exist P a h).
  Lemma dest_mk x : P x = (dest (mk x) = x).
  Proof. ... Qed.
  Lemma mk_dest x : mk (dest x) = x.
  Proof. ... Qed.
End Subtype.
\end{lstlisting}
\noindent
where \verb+{x:A|P x}+ is a notation for {\tt sig:(forall
  A:Type,(A->Prop)->Type)} with constructor {\tt exist:(forall
  A:Type,forall x:A,P x -> sig A P)} and, for all {\tt Q:Prop}, {\tt
  T:Type}, {\tt f:Q->T} and {\tt g:\tilde Q->T}, {\tt COND\_dep Q T f
  g} builds a term of type {\tt T} by using, thanks to the excluded
middle, either {\tt f} and a proof of {\tt Q}, or {\tt g} and a proof
of the negation of {\tt Q}.

Remark: to prove the equations {\tt dest\_mk} and {\tt mk\_dest}
ensuring that {\tt\{x:A|Px\}} is in bijection with the elements of
{\tt A} satisfying {\tt P}, we need to use the property of {\tt
  proof\_irrelevance} saying that any two proofs of the same
proposition are equal, to have two dependent pairs {\tt(exists P a
  h1)} and {\tt(exists P a h2)} equal even if {\tt h1} and {\tt h2}
are not definitionally equal. This does not hold in general (\eg in
homotopy type theory) but holds here because it follows from
excluded middle \cite{coquand89tr} or propositional
extensionality\footnote{\url{https://rocq-prover.org/doc/v9.0/stdlib/Stdlib.Logic.ClassicalFacts.html}}.

This default alignment can be useful, \eg for proving that HOL-Light
reals are isomorphic to Rocq reals (see below). However, in general,
we prefer to align HOL-Light types with more idiomatic Rocq types. The
following paragraphs explain how to do that, more or less
automatically, for inductive predicates, inductive types and recursive
functions.

\vspace*{1mm}\noindent{\bf Alignment of inductively defined predicates}.
An inductive predicate is defined in HOL-Light by the command
\href{https://hol-light.github.io/references/HTML/new_inductive_definition.html}{\tt
  new\_inductive\_definition} as the smallest predicate satisfying
some user-defined deduction rules.

Take for instance the \href{https://github.com/jrh13/hol-light/blob/3170739521d88d04580f61385c95b497690b7002/sets.ml#L128-L135}{finiteness} property for sets over some type $A$,
represented by their characteristic functions, \ie terms of type
$A\a\mt{bool}$. It is inductively defined in HOL-Light with the
following two rules:
\begin{itemize}
\item {\tt EMPTY A := fun \_ => False} is finite,
\item {\tt INSERT A x S := fun y => y = x $\vee$ S y} is finite if {\tt S}
  is finite.
\end{itemize}
The non-aligned translation of this definition gives in Rocq:
\begin{lstlisting}[language=Rocq]
Definition FINITE (A: Type) (S: A->Prop): Prop :=
  forall F: (A->Prop)->Prop,
    (forall T:A->Prop,
      (T = EMPTY A
      \/ exists x U, T = INSERT A x U /\ F U)
      -> F T)
    -> F S
\end{lstlisting}
which would rather be defined as follows:
\begin{lstlisting}[language=Rocq]
Inductive Finite (A: Type): (A->Prop)->Prop :=
| EMPTY_is_finite: Finite (EMPTY A)
| INSERT_is_finite:
  forall x S, Finite S -> Finite (INSERT A x S).
\end{lstlisting}
But it is possible to prove that these two definitions are in fact
extensionally equal, and we developed a tactic \href{https://github.com/Deducteam/rocq-hollight/blob/5496c6079ec6caee55db05e3cca34b482d11206a/init.v#L889-L906}{\tt ind\_align} (the code of tactics is available in the file \href{https://github.com/Deducteam/rocq-hollight/blob/main/init.v}{\tt init.v}) to
prove this kind of equalities automatically as follows:

\begin{lstlisting}[language=Rocq]
Lemma FINITE_def {A : Type'} : Finite A = FINITE A.
Proof. ind_align. Qed.
\end{lstlisting}
\noindent
The tactic first applies the extensionality axioms and then tries
proving that, for all {\tt s}, {\tt Finite A s} is equivalent to {\tt
  FINITE A s}, using the induction principles associated with each
predicate. This tactic tries solving all generated subgoals and
leaves the rest to the user.

In theory, it may fail on subgoals because the predicate's rules could
be very complex. We use a tactic based on Rocq's {\tt eauto} to try
solving them. This tactic has no guarantee of success, but it has never
failed in practice when the HOL-Light and Rocq predicates had the same rules.
Here are detailed statistics about the 33 instances of the
{\tt new\_inductive\_definition} command that we have aligned during our work:
\begin{itemize}
\item For 18 of them, the tactic allowed to prove the equality
      goal in one line.
\item For 6 others, some subgoals had to be proven by hand
      because the Rocq definition differed from the HOL-Light one.
      These were due to HOL-Light syntactic sugar producing complex
      internal definitions.
\item In 5 cases using nested instances of the {\tt Forall} or
      {\tt Forall2} predicates on lists, the proof would have been as
      easy as in the first case if Rocq had been able to generate the correct
      induction principle for the predicate as described in \cite{lamiaux25draft}.
      We instead had to prove it by hand.
\item 2 of them were the simultaneous definition of 4 mutually
      defined predicates also using the {\tt Forall} predicate.
      Mutually defined predicates are not supported by the tactic,
      so after proving the correct induction principles, we wrote specific
      tactics to handle these cases.
\item We aligned the last two with non-inductive definitions
      in Rocq, in which case the tactic cannot work as it relies
      on induction.
\end{itemize}

\vspace*{1mm}\noindent{\bf Alignment of inductive types.}
To align the subset of elements of some type $A$ satisfying some
predicate $P$ with some type $B$, we have to define some functions
$mk:A\a B$ and $dest:B\a A$ satisfying the equations showing that the
two types are in bijection. Since we have a choice function
$\varepsilon$, we can always define the inverse $f^{-1}$ as $\l
y,\varepsilon(\l x,y=fx)$. In this case:
\begin{itemize}
\item if $f$ is injective then, for all $b:B$, $f^{-1}(f(b))=b$, which
  is the first equation that must be satisfied by HOL-Light types if
  we take $f=dest$ and $mk=dest^{-1}$;
\item for all $a:A$, $a$ is in the image of $f$ iff $f(f^{-1}(a))=a$,
  which is the second equation that must be satisfied if we take
  $f=dest$ and $mk=dest^{-1}$.
\end{itemize}
\noindent
Hence, to align the two types, it is enough to define $dest:B\a A$ and
prove that it is injective and that its image is made of the elements
of $A$ satisfying $P$.

To automate those proofs when $A$ is some instance of $\mt{recspace}$,
$B$ is a Rocq inductive type and $dest$ is defined recursively,
we developed two tactics:
\begin{itemize}
\item \href{https://github.com/Deducteam/rocq-hollight/blob/5496c6079ec6caee55db05e3cca34b482d11206a/init.v#L980-L1000}{\tt \_mk\_dest\_inductive} for proving goals of the form
  $mk(dest(b))=b$. First, it applies the first lemma above to reduce
  the goal to proving that $dest$ is injective, \ie $\all b_1
  b_2,dest(b_1)=dest(b_2)\A b_1=b_2$, which it tries proving by
  induction on $b_1$ and $b_2$, using the facts that {\tt CONSTR} and
  {\tt FCONS} are injective.
\item \href{https://github.com/Deducteam/rocq-hollight/blob/5496c6079ec6caee55db05e3cca34b482d11206a/init.v#L1015-L1028}{\tt \_dest\_mk\_inductive} for proving goals of the form
  $(P~a)=(dest(mk~a)=a)$. First, it applies the second lemma above to
  reduce the goal to proving that $a$ is in the image of $dest$
  implies $P~a$, and that $P~a$ implies that $a$ is in the image of
  $dest$, \ie $\ex b,a=dest~b$. For the first subgoal, it proceeds by
  induction on $b$. For the second subgoal, it proceeds by induction
  on the definition of $P$.
\end{itemize}

\noindent
In the case of lists, it amounts to write (assuming that {\tt P} is
equal to $P_\mt{list}$):

\begin{lstlisting}[language=Rocq]
Fixpoint _dest_list {A : Type'} (l : list A) : recspace A :=
  match l with
  | nil => CONSTR 0 (ε (fun _ => True)) FNIL
  | a::l => CONSTR 1 a (FCONS (_dest_list l) FNIL)
  end.

Definition _mk_list {A} := inverse (@_dest_list A).

Lemma lemma_15 {A} (a:list A) : (_mk_list (_dest_list a)) = a.
Proof. _mk_dest_inductive. Qed.

Lemma lemma_16 {A} (r:recspace A):
  (P r) = ((_dest_list (_mk_list r)) = r).
Proof. by _dest_mk_inductive. Qed.
\end{lstlisting}
to be compared with the
\href{https://github.com/Deducteam/coq-hol-light/blob/0c8144ab3d79c5f6e9ab59205ecf14ddd57263d4/coq.v#L1417-L1468}{50
  lines} we had to write before having these tactics.

The only customizable part of a single inductive type definition in
HOL-Light is the number of constructors and, for each constructor, the number of arguments and the types of its arguments. These parameters do not impact
the structure of the proofs, so these tactics will always fully prove the goals
as long as there is a one-to-one mapping between the constructors of the two
types to align, such that a constructor and its image have the same
number of arguments of each type.

The tactics, however, are not designed to work for mutually defined
types (which exist in both HOL-Light and Rocq). Both HOL-Light and Rocq
also allow constructor arguments to have a type that is an instance
of a previously defined inductive type parametrized by a type being defined
(for example, an argument of type {\tt list A} in a constructor of the type
{\tt A}). Internally, this is dealt with by using mutually defined types,
meaning that the tactics do not support these arguments either.

We used these two tactics to align 5 inductive types in our
work out of the 6 we encountered. The only exception was the type
{\tt term} of first-order terms, which has a constructor argument of type
{\tt list term}. We were however able to adapt the proof idea to that
case (internally, two mutually defined types) by hand.

\vspace*{1mm}\noindent{\bf Conditions to align user-defined constants.}
As explained in the introduction, a HOL-Light function symbol $f$ is
defined by an axiom of the form $\th f=t$. For instance, the unaligned
definition of the HOL-Light constant {\tt \_0} representing the number
$0$ (defined in the introduction) is translated to Rocq as:

\begin{lstlisting}[language=Rocq]
Definition _0 : num := mk_num IND_0.
Lemma _0_def : _0 = mk_num IND_0.
Proof. exact (eq_refl _0). Qed.
\end{lstlisting}

For the replacement of {\tt \_0}, the zero of HOL-Light, by {\tt O},
the zero of the Rocq standard library, to not break any proof using
the lemma {\tt \_0\_def}, we need to prove that this lemma still holds
when {\tt \_0} is replaced by {\tt O}:

\begin{lstlisting}[language=Rocq]
Lemma _0_def : O = mk_num IND_0. Proof. ... Qed.
\end{lstlisting}

Once done, we can add the following entries in {\tt mapping.lp} and
redo the translation so that, this time, {\tt \_0} gets replaced by
{\tt O}:

\begin{lstlisting}[language=Lambdapi]
builtin "0" ≔ _0;
builtin "_0_def" ≔ _0_def;
\end{lstlisting}

We can proceed similarly for each user-defined constant that we want
to map to a function already defined in a Rocq library, or to a more
idiomatic definition.

Many constants are defined using $\varepsilon$ as some element
satisfying some predicate $P$. But $\varepsilon\,P$ is equal to any
other element satisfying $P$ as soon as two elements satisfying $P$
are equal, that is, if $P$ uniquely characterizes that element:

\begin{lstlisting}[language=Rocq]
Lemma align_ε (A : Type') (P : A -> Prop) a :
  P a -> (forall x, P a -> P x -> a = x) -> a = ε P.
\end{lstlisting}

\vspace*{1mm}\noindent{\bf Alignment of primitive recursive functions.}
Suppose that we want to prove in Rocq that the translation of the
HOL-Light function {\tt ADD} is equal to the function {\tt add} that
is defined in the Rocq standard library with the
\href{https://rocq-prover.org/doc/V9.2.0/refman/language/core/inductive.html#top-level-recursive-functions}{Fixpoint}
command. Remember that {\tt ADD} is defined as $\varepsilon~P~43$ for
some predicate $P$. We first change the goal {\tt add = ADD} to
{\tt(fun \_ => add)~43 = $\varepsilon$ P 43}, which is $\beta$-equivalent and,
by congruence, can be proved from {\tt(fun \_ => add) =
  $\varepsilon$ P}. After applying the lemma {\tt
  align\_$\varepsilon$} above, we are left to prove that:
\begin{itemize}
\item {\tt(fun \_ => add)} satisfies the equations defining {\tt
  ADD}. In a simple case like this, where each HOL-Light equation
  corresponds to a pattern-matching clause in the Rocq definition,
  this easily follows from the reflexivity of equality since, in Rocq,
  terms are definitionally equal modulo recursive definitions.
\item any two functions satisfying those equations are equal. To prove
  this property, we can use functional extensionality and do some
  induction on the argument on which the functions are recursively
  defined. In each generated subgoal, applying the equation satisfied
  by the function, and the induction hypothesis, turns the
  goal into an equality that can be discharged by reflexivity.
\end{itemize}
To automate this process, we defined a tactic called \href{https://github.com/Deducteam/rocq-hollight/blob/5496c6079ec6caee55db05e3cca34b482d11206a/init.v#L1144-L1229}{\tt
  total\_align} which tries to automatically prove the equality of the
HOL-Light and Rocq definitions:

\begin{lstlisting}[language=Rocq]
Lemma ADD_def : add = ADD. Proof. by total_align. Qed.
\end{lstlisting}

Our tactic can take optional parameters: the induction principle to
use (in case the one automatically generated by Rocq is not suitable,
\eg with nested inductive types \cite{lamiaux25draft}) and a solving
tactic to use in each subgoal generated by the induction (if using the
equalities from the context is not sufficient).

It works for almost all the primitive recursive functions defined with
the HOL-Light
\href{https://hol-light.github.io/references/HTML/new_recursive_definition.html}{\tt
  new\_recursive\_definition} command like the addition on Peano
numbers. The only case for which this tactic does not succeed is when
a recursive call appears under a binder because, in this case, Rocq's
{\tt rewrite} tactic fails to apply the induction hypothesis. This happened
only once, with the universal quantifier in the function
\href{https://github.com/Deducteam/rocq-hollight/blob/5496c6079ec6caee55db05e3cca34b482d11206a/Unif/mappings.v#L421-L433}{\tt
  holds} which defines the truth value of a formula.
This tactic also works independently of the Rocq function chosen for the
alignment, as it actually does not try to prove that it satisfies
the equations defining the HOL-Light function. It is the only
part which depends on the Rocq function's definition, and will either
be trivially provable without automation (like for {\tt add}) or
arbitrarily hard depending on the chosen function, as it can be anything.

Out of the 72 functions we aligned which were defined with
the {\tt new\_recursive\_definition} command, the {\tt total\_align}
tactic worked in 71 cases (all but {\tt holds}),
though some goals needed more work due to the Rocq function we chose
for the alignment. 


\section{Alignment of the type of real numbers}
\label{sec-trans-real}
In this section we describe how we aligned the type of real numbers
from HOL-Light with the one in Rocq
\href{https://rocq-prover.org/doc/V9.1.0/stdlib/Stdlib.Reals.Rbase.html}{standard}
library. We start by explaining how real numbers are defined in both
systems.

\vspace*{1mm}\noindent{\bf Real numbers in HOL-Light.}
The HOL-Light file
\href{https://github.com/jrh13/hol-light/blob/master/realax.ml}{realax.ml}
axiomatizes the type $\mt{real}$ of real numbers from natural numbers
in several steps of subtype and quotient type constructions
\cite{harrison98book}, a quotient type being a particular case of
subtype: given an already defined type constant $A$ and an equivalence
relation $R:A\a A\a\mt{bool}$, the quotient type ${B}={A/R}$ is
axiomatized as an isomorphic image of the subset of the elements of
$A\a\mt{bool}$ that are equivalence classes wrt $R$, \ie that
satisfies the predicate ${P(x)}={(\ex y,x=Ry)}$:
\begin{enumerate}
\item HOL-Light first defines the type $\mt{nadd}$ as an isomorphic
  image of the subset of $\mt{num}\a\mt{num}$ of the functions $f$ on
  natural numbers that are nearly additive, that is, such that $\ex
  B,\all m,\all n, |m\times f(n)- n\times f(m)|\le B\times(m+n)$,
  which implies that the sequence $(x_n)_{n\ge 1}$ with $x_n=f(n)/n$
  is a Cauchy sequence (sequence whose elements become arbitrarily
  close to each other as the sequence progresses). It then defines
  various operations and predicates on $\mt{nadd}$: $\le$, $+$, $-$,
  $\times$, $/$, \ldots

\item HOL-Light then defines the type $\mt{hreal}$ of non-negative
  reals as an isomorphic image of the quotient of $\mt{nadd}$ by the
  relation ${R\,x\,y}={\ex B,\all n,|x_n-y_n|\le B}$. The operations
  and predicates on $\mt{nadd}$ are then lifted to $\mt{hreal}$.
  
\item Finally, it defines the type $\mt{real}$ as an isomorphic image
  of the quotient of $\mt{prod}~\mt{hreal}~\mt{hreal}$ by the
  equivalence relation $R$ such that ${R\,(x_1,y_1)\,(x_2,y_2)}$ =
  ${(x_1+_\mt{hreal}y_2=x_2+_\mt{hreal}y_1)}$, called $\mt{treal\_eq}$
  in HOL-Light.
\end{enumerate}

\noindent{\bf Real numbers in Rocq.}
There are various axiomatizations or constructions of real numbers in
Rocq \cite{boldo16mscs}. We only discuss some of them hereafter.

The \href{https://github.com/coq-community/fourcolor}{Fourcolor}
library provides a construction of real numbers using Dedekind cuts
and the axiom of excluded middle only \cite{gonthier05tr}.

The Rocq standard library was initially declaring a type $\mt{R}$ with
some operations and axioms stating that $\mt{R}$ is a
Dedekind-complete Archimedian\footnote{This axiom is actually
redundant as it follows from Dedekind-completeness.} ordered field
\cite{mayero01phd}. Later, Séméria restructured the Rocq standard
library by first defining an axiomatization and an implementation of
constructive reals using no axiom, and then an axiomatization and
implementation of classical reals on top of constructive reals using a
few axioms\hide{\footnote{(Dependent) functional extensionality ($(\all
x,fx=gx)\A f=g$), excluded middle for negated propositions ($\all
P:Prop,\neg P\ou \neg\neg P$), and limited principle of omniscience
(LPO) saying that, for any sequence of propositions $P(n):Prop$, one
can decide whether $\all n,P(n)$ or $\ex n,\neg P(n)$ holds, assuming
that, for all $n$, one can decide whether $P(n)$ or $\neg P(n)$
holds.}} satisfied by a set-theoretical model of the calculus of
inductive constructions \cite{semeria20jfla}.

\vspace*{1mm}\noindent{\bf Alignment of real numbers with the Rocq standard library.}
By mapping every HOL-Light type used in the construction of real
numbers to the corresponding default subtype construction in Rocq, we
can translate the HOL-Light construction of real numbers to Rocq
without adding new axioms other than those already used in HOL-Light.
We are then left to prove that the obtained type is in bijection with
the type {\tt R} defined in the Rocq standard library, which follows
from the fact that any two Dedekind-complete ordered fields are isomorphic
\cite{huntington03tams}, a result formalized in Rocq both in the
C-CoRN library for constructive reals \cite{geuvers00types-reals} and
in the
\href{https://github.com/rocq-community/fourcolor/blob/master/theories/reals/realcategorical.v}{Fourcolor}
library for classical reals \cite{gonthier05tr}.

As HOL-Light is based on classical logic, we used the Fourcolor
library, in which a model of real numbers is defined as the
combination of two records\footnote{A record maps a fixed number of
field names to some
values.}\footnote{\url{https://github.com/rocq-community/fourcolor/blob/master/theories/reals/real.v}}:
a first record providing a type, a relation $\le$ and some functions
on it (addition, subtraction, multiplication, division, supremum,
\ldots), and a second record providing proofs of the properties that
these functions should satisfy. \hide{It is actually a setoid model
  \cite{hofmann95tlca} as $\le$ is not required to be anti-symmetric.}
We then proved that both the type {\tt R} defined in the Rocq standard
library and the type obtained by translating the HOL-Light type {\tt
  real} to Rocq are models of reals and thus isomorphic.

\hide{We also proved that, in both models, the equivalence relation
  associated with $\le$ is in fact equal to the propositional equality
  {\tt Logic.eq} defined in the Rocq standard library.}
Afterwards, we defined a function {\tt mk\_real} from the type of
subsets of {\tt prod hreal hreal} to the type {\tt R} of real numbers
of the Rocq standard library, and its inverse {\tt dest\_real}, and
proved that they satisfy the required equations.

The fact that we have an isomorphism between {\tt real} and {\tt R}
and not just a bijection is essential to later align the basic
operations and predicates on reals (addition, multiplication, \ldots)
which are defined by lifting to equivalence classes the operations on
{\tt prod hreal hreal}.


\section{Translation of HOL-Light libraries to Rocq}

\vspace*{1mm}\noindent{\bf HOL-Light base library.}
The methods and tactics described above have been used to align the
following data types and functions of the HOL-Light base
library
\href{https://github.com/jrh13/hol-light/blob/master/hol_lib.ml}{\tt
  hol\_lib} with both those of the Rocq standard library and those of
the Rocq Mathcomp library (\cite{blanqui24lpar} only aligned
propositions, the unit type, the product type and the type of natural
numbers in base 1):
\begin{itemize}
\item propositions: connectives and their intro/elim rules, equality, $\varepsilon$
\item relations: well-foundedness and accessibility predicates
\item unit type: its unique constructor
\item sum type: its constructors
\item option type: its constructors
\item product type: constructor and projections
\item natural numbers (both in base 1 and in base 2): $0$, successor, doubling function, predecessor, $+$, $\times$, exponentiation, $\le$, $<$, $\ge$, $>$, $\min$, $\max$, factorial, Euclidian division $/$, modulo, even and odd predicates
\item lists: constructors, concatenation, reversal, length, map, map2, last element removal, forall/exists predicates, filter, membership, element repetition, iterator, head, tail, last element, $n$-th element, zip, fold, \ldots
\item reals: $\le$, $<$, $\ge$, $>$, $+$, $-$, $\times$, $/$, inverse $.^{-1}$, opposite, exponentiation, $|~|$, max, min, sign, square root $\sqrt.$, modulo
\item integers (both in base 1 and in base 2)\footnote{In HOL-Light, integers are defined as a subset of reals while, in Rocq, there are inductively defined as signed numbers in base 1 (Mathcomp) or 2 (standard library).}: injection of natural numbers, $\le$, $<$, $\ge$, $>$, $+$, $-$, absolute value $|~|$, sign, max, min, exponentiation, Euclidian division, modulo, remainder, co-primality, gcd, lcm
\item sets: $\emptyset$, $\in$, $\cup$, $\cap$, image, difference, singleton, disjointness
\item finite sets: finiteness predicate, cardinal, iterator, conversion to/from lists
\end{itemize}

\vspace*{1mm}\noindent{\bf HOL-Light Logic library.}
The alignment tactics helped us also in generating an idiomatic
translation in Rocq of the HOL-Light
\href{https://github.com/jrh13/hol-light/blob/master/Logic/make.ml}{Logic}
library.  We replaced all HOL-Light definitions by idiomatic Rocq
definitions using the {\tt Inductive} and {\tt Fixpoint} commands with
our tactics and the {\tt Equation} command \cite{sozeau19icfp} for
functions defined by well-founded induction rather than structural
recursion.  We aligned the types of first-order terms and formulas,
and the recursive functions or predicates defined on them, including
prenex normalization, skolemization, unification, multiple algorithms
for resolution, the lexicographic path ordering.

This library features a huge proportion of definitions for which one
of our tactics applies, and we spent more than 50\% of our time doing
such easy alignments, so the tactics were really helpful, definitely
saving us more hours than we spent writing and documenting
them. Although proofs using tactics can sometimes be hard to maintain,
in our experience, they helped us a lot as one change can solve an
issue in dozens of similar goals. But our tactics apply to very
specific cases, always with a recognizable pattern, and they follow a
clear proof scheme, so we did not have maintenance issues.

The obtained library contains, among other results, proofs of the
Skolem-Gödel-Herbrand, Herbrand and Birkhoff theorems, and correctness
and completeness proofs of several resolution algorithms which, to the
best of our knowledge, had not been formalized in Rocq before.

\vspace*{1mm}\noindent{\bf HOL-Light Multivariate and Rocq Mathcomp-analysis.}
Now that the HOL-Light definition of reals has been aligned with the
one of Rocq, we can turn our attention to the translation and reuse in
Rocq of results that are already formalized in the HOL-Light
\href{https://github.com/jrh13/hol-light/blob/master/Multivariate/make_complex.ml}{Multivariate}
library, which contains more than 20,000 theorems on real and complex
analysis. On the Rocq side, we will use the
\href{https://github.com/math-comp/analysis}{\tt Mathcomp-analysis}
library \cite{affeldt24itp}.

We describe hereafter what we did to get a directly usable translation
of an interesting theorem of Multivariate that was not available
in Rocq before:
\href{https://github.com/jrh13/hol-light/blob/3170739521d88d04580f61385c95b497690b7002/Multivariate/topology.ml#L22843-L22929}{\tt
  SERIES\_RATIO}, the ratio test for the convergence of series on
$\bR^{N'}$, saying that, for every integer $\mt{N'\ge 1}$, real
$\mt{c}<1$, sequence $\mt{a}$ on $\bR^{N'}$, subset
$\mt{s}\subseteq\bN$, and integer $N''\ge 0$, if, for all $n\ge N''$,
$||a_{n+1}||\le c\,||a_n||$, then there is $l$ such that
$\displaystyle{\lim_{n\a\infty}\Sigma_{i\in\mt{s}\cap[0,n]}\,a_i=l}$.
\begin{itemize}
\item For some basic types, Mathcomp has its own type definitions (\eg
  reals and integers), and for some others ({\tt nat}, {\tt list}), it
  has its own implementation of all the basic definitions to reduce
  its dependencies to the Rocq standard library. Therefore, we had to
  update the mappings we previously developed for these types with the
  Rocq standard library.

\item Since HOL-Light has no dependent types, it cannot define the
  cartesian power $\mt{cart}~\al~n$ with $n:{\tt num}$. Instead, it
  uses $\mt{cart}~\alpha~\eta$ where $\eta$ is a type of cardinality
  $n$ \cite{harrison05tphol}. On the other hand, Mathcomp-analysis
  uses the dependent type $\mt{`rV[\alpha]}_n$ of row matrices of size
  $n$. We however proved that these two types are equal:

  \begin{lstlisting}[language=Rocq]
Lemma row_to_cart (A:Type') (n:nat) (gt0n: 0<n):
  'rV[A]_n = cart A (@enum_type n gt0n).
  \end{lstlisting}

  where {\tt(@enum\_type n gt0n)} is the subtype of all naturals
  smaller than {\tt n}.
\item Although HOL-Light defines the general notion of limit through
  metric nets and Mathcomp-analysis through filters, the derived
  definition of limit for $\mathbb{R}^n$ sequences are identical in both
  systems (standard $\all\vep\ex\eta$ definition).

\item HOL-Light uses the Euclidean norm for vectors, whereas
  Mathcomp-analysis uses the infinite/sup norm, without support for
  other norms. So we had to prove some lemmas not present in
  Mathcomp-analysis stating the equivalence of these two norms with
  their optimal bounds.
\end{itemize}
Then, we aligned the following notions: sum over a set $\Sigma_{x\in
  S}f(x)$ (defined as $0$ if $\{x\!\in\!S\mid\!f(x)\!\neq\!0\}$ is
infinite), vector coordinates, vector subtraction, scalar product (which we
defined using Mathcomp-analysis linear forms), Euclidean norm (which
we defined using the Mathcomp square root function) and the resulting
distance, metric nets (our own definition), convergence in a metric
net, and finally the convergence of a vector series over a set. The
proofs take about 500 lines (see the file \href{https://github.com/Deducteam/rocq-hollight/blob/main/Multivariate/mappings.v}{\tt mappings.v}).

In the end, we obtain after unfolding some definitions the following
theorem:

\begin{lstlisting}[language=Rocq] Lemma
thm_SERIES_RATIO: forall {N':Type'} (c:R) (a:nat->cart R N')
  (s:nat->Prop) (N'':nat), c<1 /\ (forall n:nat, (N''<=n)%N
    -> vector_norm (a n.+1) <= c * vector_norm (a n)) ->
  exists l:cart R N', \sum_(i<n.+1|s i) a i @[n-->\oo] --> l.
\end{lstlisting}
\noindent
where notations for comparisons, product, limit and sum come from
Mathcomp itself, and $\mt{vector\_norm}$ is the Euclidean norm we had
to define.

Note that, here, $\mt{N'}$ is a type and not an integer, and that {\tt
  cart R N'} is $\bR^{|\mt{N'}|}$ where $|\mt{N'}|$, the cardinal of
$\mt{N'}$, is $1$ if $\mt{N'}$ is infinite. The theorem can however be
applied to Mathcomp developments by first using the lemma \href{https://github.com/Deducteam/rocq-hollight/blob/5496c6079ec6caee55db05e3cca34b482d11206a/HOL/mappings.v#L1282-L1288}{\tt
  row\_to\_cart} above. It should be possible to get a statement where
$\mt{N'}$ is an integer by applying some transfer tool like Trocq
\cite{cohen24esop}.

The fact that we need to unfold some definitions to get more user-friendly
statements may create some difficulties when using search tools or
tactics based on pattern matching (\eg {\tt Search} command and {\tt
  rewrite} tactic). We are working on improving this by using
meta-programming techniques or AI tools.

Another limitation of the current translation is that it can only
replace identifiers and not expressions. But, for instance, HOL-Light
has no identifier for the types of matrices and complex
numbers. Instead, it uses the expressions $(\bR^n)^m$ and $\bR^2$
respectively. This problem can however be solved by modifying
HOL-Light sources beforehand, or by using transfer tools afterwards.

\section{Related work}
\label{sec-related}

In various proof assistants, transfer libraries and tactics have been
developed
\cite{sacerdoti04types,sozeau09jfr,huffman13cpp,cauderlier17itp,kappelmann23aplas,cohen24esop}
to replace a statement built from some types and functions into a
statement where the previous types and functions have been
replaced by new ones, given that the new types are
isomorphic to the former ones, and that the new functions are
morphisms between those types, \ie mapping related inputs to related
outputs. We do not need to use them here because we do not use
arbitrary equivalence relations but simply replace equals by equals,
and can thus do the replacements during the translation itself. If we
were doing no replacement during the translation, we would then have
to use those transfer tactics in order to get the expected
theorems. Doing the replacements during the translation is both
simpler and faster.

The other works on the translation of proofs from HOL-Light to Rocq
are Denney in 2000 \cite{denney00tphol}, Wiedijk in 2007
\cite{wiedijk07note} and Keller and Werner in 2010 \cite{keller10itp}.
In \cite{denney00tphol}, Denney presents a prototype translating a
subset of HOL-Light proof trees represented in some defined proof
format (similar to OpenTheory \cite{hurd11nfm}) to Rocq proof scripts, each
HOL-Light proof step being translated to a tactic call.
Following the approach of \cite{naumov01tphol} for translating HOL98
to Nuprl, Wiedijk showed in \cite{wiedijk07note} how one could encode
HOL-Light proofs in Rocq more efficiently, by translating each
HOL-Light proof step into a Rocq definition, approach that we
implemented in \cite{blanqui24lpar}.
Keller and Werner in \cite{keller10itp} instead use a deep
embedding of HOL-Light proofs in Rocq, and define and prove the
correctness of a boolean function checking the correctness of an
HOL-Light proof, but this software is not maintained anymore.

There is no mention of the problem of alignments in
\cite{denney00tphol}.  For getting usable theorems, in
\cite{wiedijk07note}, Wiedijk suggests to use transfer
techniques. Only Keller and Werner mention in \cite{keller10itp} that
some HOL-Light symbols can be replaced by already defined Rocq symbols
but the proofs are left to the user \cite[Section
  11.4.2]{keller13phd}.
Some alignments\footnote{File
\href{https://gitlab.inria.fr/hol-light-isabelle/import/-/blob/master/isabelle/HOL_Light_Maps.thy?ref_type=heads}{\tt
  HOL\_Light\_Maps.thy} in
\url{https://gitlab.inria.fr/hol-light-isabelle/import}} are also used
in \cite{obua06ijcar,kaliszyk13itp,bergeron24isabelle} to translate
HOL-Light libraries to Isabelle/HOL, but these two systems are very
similar and the proofs of the correctness of these mappings much
simpler.

Finally, in \cite{cauderlier17itp}, the authors provide a library of
transfer theorems for arithmetic in FoCaLiZe, an environment for
developing certified programs and libraries on top of Rocq, and use it
to check in Dedukti the correctness of a proof built using a lemma
proved in Rocq and a lemma proved in HOL-Light.


\section{Conclusion and future work}
\label{sec-conclu}



Thanks to the alignments and their formal proofs described above,
which take about 8,000 lines of Rocq, one can now automatically
translate HOL-Light theorems on HOL-Light reals to Rocq theorems on
Rocq reals, and similarly for several other data types and functions
on those data types (natural numbers, integers, lists, \ldots), by
using the HOL-Light to Dedukti translator
\href{https://github.com/Deducteam/hol2dk}{hol2dk} and the Dedukti to
Rocq translator \href{https://github.com/Deducteam/hol2dk}{lambdapi}
\cite{blanqui24lpar}.

As an application, we translated to Rocq the whole HOL-Light Logic and
Multivariate libraries which contain more than 20,000 theorems on
arithmetic, logic, real and complex analysis, \ldots\ hence providing to
Rocq users theorems that had not been formalized in Rocq before (\eg
the ratio test for the convergence of series discussed above), and
that are now directly usable since they talk about types that are
defined in the Rocq standard or Mathcomp libraries.

Obviously, many of these 20,000 theorems will be redundant with
existing Rocq results, as they only describe basic properties
of the objects we aligned. The exact amount is hard to quantify
(it would be possible to use an LLM to do it, but the result would not
be certified). Note that the theorems which are not already present
in Rocq are often the ones that are the hardest to prove.
Importantly, due to the size of the library, there surely are
definitions not present in either the Rocq standard
or Mathcomp analysis libraries that users could be interested in.
This translation provides them with these objects, and despite perhaps
not being defined the way it would naturally be in Rocq, it also provides
basic properties which are often all that is needed to work with them.
Even for these, existing alignments are crucial as the translation therefore
contains theorems in which these new definitions and pre-existing Rocq
ones interact.

The proofs of the alignments and the translated definitions and
theorems are available in the Opam package
\href{https://github.com/Deducteam/rocq-hollight}{rocq-hollight}. To
enable a fast installation and import, only the definitions and
statements are provided due to the size of the proofs.
For the sake of reproducibility, we however provide a Shell script to
generate and check those proofs again. The size of those proofs could be
reduced by instrumenting the HOL-Light code more deeply so as to replace some
subproofs by tactic calls in Rocq \cite{blanqui24lpar}.

We mentioned that we could now check proofs in 10 minutes for the
HOL-Light base library and in 21 hours for the Multivariate library,
but by adding Mathcomp libraries as a dependency, proof-checking
is now much longer (yet still doable). This is due to
every generated file having to import the Mathcomp dependencies,
taking more than 7 seconds, which is particularly long since most files
used to check in less than a second. The HOL-Light base library now takes
50 minutes to check while the Multivariate library takes several days.
To solve this issue, we are currently experimenting with ways to
implement alignments after the proofs were checked, since we can then
import Mathcomp only for alignments and not for proofs. Our attempts
involve parametrizing every proof by the set of Rocq functions we wish
to align the HOL-Light ones with, and the set of proofs of their
alignments.

On another topic, more directly applicable theorems can be obtained by
proving new alignments of more advanced mathematical structures.
As explained in this paper, the alignment of two constant definitions
requires to prove one equality, and the alignment of two type definitions
requires to define one function and prove two equalities stating that this
function is a bijection. Therefore, this work can be easily extended
further independently, all the more so since we developed several
tactics to prove those equalities automatically.

Finally, one may wonder to which extent the current work can be
extended to other systems, either by changing the target system
(Rocq), or by changing the source system (HOL-Light). The Dedukti to
Rocq translator for definitions and proofs in higher-order logic could
be extended to other systems similar to Rocq like Matita, Agda
\cite{felicissimo24lmcs} or Lean \cite{blanqui26ictac}. But, because
all the alignments currently used are proved in Rocq, they cannot be
directly reused within other systems: for each target system, we need
to define and prove the correctness of all alignments again. However,
it might be possible to factorize some of these definitions and proofs
by moving them to Dedukti itself, where it is possible to represent
and use inductive types as they are defined in those systems. We would
then get all these mappings automatically as soon as Dedukti can be
translated to a new system. The difficulty is that Dedukti does not
feature proof tactics as powerful as the ones available in Rocq (\eg
for linear arithmetic on integers or reals) but this may change in the
future as several automated theorem provers and SMT solvers can now
output Dedukti proofs directly
\cite{coltellacci24smt,coltellacci25frocos,sutcliffe25flairs,petkovic25draft,taprogge26lpar}.

As for changing the source system (HOL-Light), this requires to
instrument the new source system so as to generate proofs in the
Dedukti format, or in the input format used by Hol2dk. For systems
having more features than HOL-Light, like Isabelle/HOL which has type
classes and locales \cite{wenzel97tphol}, this is nontrivial and
the subject of some ongoing work (see \cite{wenzel24isabelle} and
\url{https://github.com/Deducteam/isabelle_dedukti}). For systems
similar to HOL-Light, like HOL4, this should be simpler. This however
requires to change the input format of Hol2dk into a more portable
format like the ones developed to translate HOL-Light proofs to
Isabelle/HOL \cite{kaliszyk13itp,adams15pxtp}.

\hide{
The concept of alignment can easily be integrated in a translation as
long as the source system has a way of handling definitions that is similar
to HOL systems (with propositional axioms). This is not the case of systems
like Rocq, where definitions are handled through reductions that are not
expressible in the language (one cannot define a predicate equivalent to "x is defined as y",
whatever y is).
}

Finally, doing the reverse translation, from Rocq to HOL-Light (or
similar systems), is significantly harder for multiple reasons: the
term
\href{https://rocq-prover.org/doc/master/refman/language/core/conversion.html}{conversions}
that are implicit in Rocq must be made explicit in HOL-Light
\cite{blot24fossacs}, and type universes and dependent types must be
eliminated which is an open problem in general even though it is
possible in some cases \cite{thire18lfmtp,thire20phd}.


\label{sect:bib}
\bibliographystyle{plain}
\bibliography{main}



\end{document}